# Analysis of Anonymous User Interaction Relationships and Prediction of Advertising Feedback Based on Graph Neural Network

Yanjun Dai[*]

Brandeis University, 415 South St, Waltham, MA 02453, yanjundai0000@gmail.com

Haoyang Feng

Duke University, Durham, NC 27708, USA, brianmaga2024@gmail.com

Yuan Gao

Boston University, Boston, MA 02215, USA, xyan56379@gmail.com

**Abstract**

While online advertising is highly dependent on implicit interaction networks of anonymous users for engagement inference, and for the selection and optimization of delivery strategies, existing graph models seldom can capture the multi-scale temporal, semantic and higher-order dependency features of these interaction networks, thus it's hard to describe the complicated patterns of the anonymous behavior. In this paper, we propose Decoupled Temporal-Hierarchical Graph Neural Network (DTH-GNN), which achieves three main contributions. Above all, we introduce temporal edge decomposition, which divides each interaction into three types of channels: short-term burst, diurnal cycle and long-range memory, and conducts feature extraction using the convolution kernel of parallel dilated residuals; Furthermore, our model builds a hierarchical heterogeneous aggregation, where user-user, user-advertisement, advertisement-advertisement subgraphs are combined through the meta-path conditional Transformer encoder, where the noise structure is dynamically tamped down via the synergy of cross-channel self-attention and gating relationship selector. Thirdly, the contrast regularity of feedback perception is formulated, the consistency of various time slices is maximized, the entropy of control exposure information with dual-view target is maximized, the global prototype of dual-momentum queue distillation is presented, and the strategy gradient layer with light weight is combined with delaying transformation signal to fine-tune the node representation for benefit-oriented. The AUC of DTH-GNN improved by 8.2% and the logarithmic loss improved by 5.7% in comparison with the best baseline model.

**CCS CONCEPTS**

• Computing methodologies ~ Machine learning ~ Machine learning approaches ~ Neural networks

## 1 INTRODUCTION

In the domain of online advertising, the anonymity and dynamic of user behavior is one of the key challenges for the current ad feedback modeling. The ad content is commonly browsed by the user from different non-registered/multi-platform/non-identifiable access behaviors, the anonymous behaviors without identifier exchanged between interactions do not act as the stable identifier properties, and we do not have a way to archive and model the user's historical behavior for user behavior. Meanwhile, advertising interaction is sparse, unstructured and real-time, which makes it even harder for the model to catch the interests of real users [1]. Thus, in view of the fact that we cannot simply trust the personally identifiable information of users, how to mine useful information from fragmentary behavior trajectory and how to build a feasible advertising feedback prediction mechanism with generalization ability are both the crucial breaking points in the current research about recommender system.

---

[*] Corresponding author.

Traditional recommender models, including Collaborative Filtering and Matrix Factorization, need to know explicit user IDs and the whole click logs for modeling, while such two recommender systems based on complete click logs usually issues privacy issues. They do not perform well using anonymized user information, but it is difficult for them to make better use of possible structural or contextual semantic information among the users [2]. Inspite of the fact that high-order features could be excavated by deep learning models, a data without identifying information can result in overfitting or poor generalization. So there is an urgent need for a modeling approach with explicit topology description of the interactive network and good robustness and information propagation performance [3].

To address the aforementioned issues, several structural improvement strategies and context-aware mechanisms are suggested in this paper. Prior works have demonstrated the efficacy of combining transfer learning and lightweight CNN backbones such as MobileNetV2 in extracting structured representations from image-based medical data, which inspires the feasibility of analogous modular adaptation for anonymous behavior graphs [23]. First, the Fi-GNN model proposes the Feature Interaction Graph Construction, explicitly treating advertising context factors (e.g., location information, device type, timestamp, etc.) as edges in a graph, which could benefit the model learning to capture nonlinear cross-field dependencies [4]. In contrast, SAN-GNN utilizes a time-sensitive adjacency aggregation scheme which incorporates the interaction time into the message passing process as part of the edge right and enhances the model's capability of capturing users' short-term interest. Such results indicate the flexibility of GNNs in modeling complex behavior tasks and also justify the theoretical foundation of anonymous feedback modeling [5].

While the current literature has achieved notable progress in terms of the model structure and the privacy mechanism, recent research further explores the application of AI in fields like credit risk detection [17] and ad placement optimization in e-commerce using ensemble learning and predictive modeling frameworks [18]. There still exist some major challenges for the prediction task anonymous user advertisement feedback. User interaction data is usually heterogeneous due to different sources, such as user behavior, content features, and contextual environment, but the current GNN models lack the homogeneous and efficient fusion mechanism of heterogeneous data [6]. Significantly, advertising feedback signals usually have delay features, say clicks, conversions, payments, etc., which makes more strict requirement on the long-term dependence modeling ability of the model, similar to how time-series learning frameworks such as CNN-LSTM or attention-based mechanisms are leveraged in meteorological and activity recognition contexts [13,14,15]. Finally, the feedback signals are often sparse, biased, and even dirty, and if the model fails to introduce the adversarial mechanism or contrastive learning strategy well in the training process, it is likely to fall into local optima [7], especially when facing highly imbalanced feedback labels, a challenge commonly seen in fraud detection and now mirrored in ad response modeling [22]. Therefore, to meet the challenge, the research issue of anonymous advertising prediction is to develop a Graph Neural Network

(GNN) framework with structural modeling capability, timing modeling capability, and feedback self-calibration capability.

## 2 RELATED WORK

Ciuchita et al. [8] suggested three affordances: interactivity, visibility, and anonymity. Such functions give users more ways to express themselves and get involved in digital platforms, yet also bring new problems for the platforms themselves, for example, how to handle the anonymous feedback and how to ensure the interaction quality on the platform. Wu et al. [9] proposed a federated learning based approach named FedCTR to predict CTRs for native ad. We show that our approach can learn user interest representations from user behavior across different platforms with user privacy preserved. FedCTR applies a local user model on each platform to make local user embedding from local user behavior, and then uploads the local user embedding to a server for aggregation and CTR prediction.

Ahn et al. [10] emphasize the complicated interactions of advertisers, platforms and users, particularly in virtual worlds where these relations become more varied and lively. According to the research, anonymity and interactivity with metaverse advertising play a crucial role on both user engagement and advertising effectiveness. The framework offers a new view of and a methodological ground on which a future virtual advertising research can be done. Lopes et al. [11] overset the emphasis of content marketing in appealing customers, establish trust in brand and increase long-term sales. The research explains that desipite being well used in practice, the effectiveness of content marketing still remains an under-researched topic in the academic world.

Zhang et al. [12] presented a new model named as CoHHN: Collaborative Guided Heterogeneous Supergraph Network for addressing the issue of disregarding price consideration in session recommendation systems. The classic session recommendation approach is to recommend according to users preferences and interests, however, we ignore the most critical characteristic of price, which will be the main factor for users in making purchase decisions. The CoHHN model fuses multiple feature information of commodities based on the formation of heterogeneous hypermaps and proposes a dual-channel aggregation mechanism to excavate user's price preference and interest preference independently. De Cosmo et al. [13] analysed user attitude to chatbots and privacy concern as being part of the relationship between mobile advertising attitude and intensity of chatbot use. Through the development of a moderation mediation model, it is demonstrated that attitudes toward mobile advertising do not directly influence behavioral intentions to use chatbots but critically impact attitudes toward chatbots. Moreover, privacy concerns negatively moderate the relationship, i.e., the higher the level of privacy concerns, the weaker the positive influence of users' positive attitudes toward chatbots on their use intentions.

## 3 METHODOLOGIES

### 3.1 Temporal Edge Decomposition and Residual Convolution

In order to solve the problem that advertising interaction has the multi-scale characteristics of "second-level sudden day-night fluctuation week/month inertia", DTH-GNN first divides the edges $(i, j, t, f_{ij})$ with timestamps into three channels according to the time difference $\Delta t_{ij} = t_{now} - t$, which is short-time (S), diurnal (D), and long-range (L), as shown in Equation 1:

$$w_{ij}^{(c)} = \frac{\exp(-\frac{\Delta t_{ij}}{\tau_c})}{\sum_{c' \in \{S,D,L\}} \exp(-\frac{\Delta t_{ij}}{\tau_{c'}})}, \quad \tau_S < \tau_D < \tau_L, \tag{1}$$

The learnable decay constant of $\tau_c$ allows the model to adaptively extract the most appropriate time granularity during training, ensuring stable modeling when bursts of traffic and inertial clicks coexist. Then follow Equations 2 and 3:

$$\tilde{f}_{ij}^{(c)} = w_{ij}^{(c)} \odot f_{ij}, \tag{2}$$

$$\tilde{h}_i^{(0)} = MLP(x_i). \tag{3}$$

Through the above formulas, the channelized edge features are constructed and the node representations are initialized to provide an information source for multi-scale convolution.

Within each layer $l$, the convolutional kernel $\kappa_r^{(c,l)}$ of the expansion rate $r \in \{1,2,4\}$ is set in parallel for each time channel to form a pyramidal receptive field, as in Equations 4 and 5:

$$m_i^{(c,l)} = \sum_r \sum_{j \in \mathcal{N}_i} \kappa_r^{(c,l)} \left(\tilde{f}_{ij}^{(c)}\right) \odot \tilde{h}_j^{(l-1)}, \tag{4}$$

$$h_i^{(c,l)} = \sigma\left(BN\left(m_i^{(c,l)}\right) + \tilde{h}_i^{(l-1)}\right). \tag{5}$$

The residual jumper not only prevents the disappearance of the deep gradient, but also retains the local and global context. Batch normalization with ReLU/Swish activation increases convergence speed and reduces internal covariate shift.

For multiple types of dependencies such as "user-ad-user" and "ad-ad", construct metapath conditions such as Transformer, as shown in Equations 6 and 7:

$$\alpha_{ij}^{\phi} = softmax\left(\frac{(q_i + p_\pi)^\top k_j^{\phi}}{\sqrt{d}}\right), \tag{6}$$

$$z_i^{\phi} = \sum_{j \in \mathcal{N}_\phi(i)} \alpha_{ij}^{\phi} v_j^{\phi}, \tag{7}$$

where $p_\pi$ is injected into the path sequential position encoding, explicitly aligning cross-type temporal relationships.

This is followed by cross-channel-cross-semantic gating, as in Equations 8 and 9:

$$\beta_{c,\phi} = \sigma\left(u^\top \tanh\left(W_1 \tilde{z}_i^{(c)} + W_2 \tilde{z}_i^{(\phi)}\right)\right), \tag{8}$$

$$h_i^{(l)} = \sum_{c,\phi} \beta_{c,\phi} z_i^{(c,\phi)}, \tag{9}$$

Dynamically suppress noise relationships and highlight critical links; Visualization $\beta$ can reveal traffic cycles such as lunch break and evening peak hours, providing explainable insights for business.

### 3.2 Feedback-Aware Contrastive Regularization

Construct a positive view $(\mathcal{G}_t, \mathcal{G}_{t+\delta})$ and a negative view $(\mathcal{G}_t, \mathcal{G}_{cf})$, corresponding to true repeated exposure and control unexposed, as shown in Equation 10:

$$\mathcal{L}_{CL} = -\log \frac{\exp\left(\frac{sim(h_u^t, h_u^{t+\delta})}{\tau}\right)}{\exp\left(\frac{sim(h_u^t, h_u^{t+\delta})}{\tau}\right) + \exp\left(\frac{sim(h_u^t, h_u^{cf})}{\tau}\right)}. \tag{10}$$

Maximize the consistency with users across time and minimize the consistency of control exposure can significantly alleviate the problem of cold-start advertising and exposure bias. This also resonates with the growing adoption of contrastive or adversarial structures in privacy-sensitive prediction tasks, such as real-time psychological state prediction in human-computer interaction [16].

Maintain the fast update encoder $\theta_q$, the slow momentum encoder $\theta_k$, and the prototype queue $\mathcal{P}$, as in Equations 11 and 12:

$$\theta_k \leftarrow \lambda\theta_k + (1-\lambda)\theta_q, \tag{11}$$

$$\mathcal{P} \leftarrow enqueue(mean(h_i)). \tag{12}$$

Slow parameters provide stable negative samples, and global prototypes capture cross-batch clustering structures; The synergy between the two can prevent feature collapse in mini-batch contrast learning and improve the embedding quality of long-tail nodes.

We formally define the strategy gradient layer as a lightweight reinforcement-based optimization module that integrates the delayed reward into node embedding updates via the REINFORCE rule, as Equation 13. Here, the delayed transformation signal refers to the observed lag between user interaction and final conversion feedback (e.g., purchases), which is used as part of the reward function. Incorporate delayed conversion revenue $r_{u,a}^{(t)}$ directly into the optimization, using REINFORCE, as in Equation 13:

$$\nabla_\theta \mathcal{J} = \mathbb{E}_{\pi_\theta}\left[\left(r_{u,a}^{(t)} - \hat{b}\right)\nabla_\theta log\pi_\theta\left(a\big|h_u^{(T)}\right)\right], \tag{13}$$

Among them, baseline $\hat{b}$ uses the moving average to reduce the variance; This layer adds only $\approx 0.3ms$ inference latency, but delivers a significant 3% increase in total $ROAS + 3\%$ in industrial $A/B$ testing.

Fig.1 shows the architecture of the "hierarchical heterogeneous aggregation module" in DTH-GNN model, in which the main idea is to explore the semantic correlations among diverse node types by building arbitrary semantic subgraphs (e.g., users -users, users -ads, ads-ads). A Transformer-based Encoder is used to process each kind of subgraph in the graph via a disparate path, and includes two major components: a gated relation selector and cross-channel self-attention.

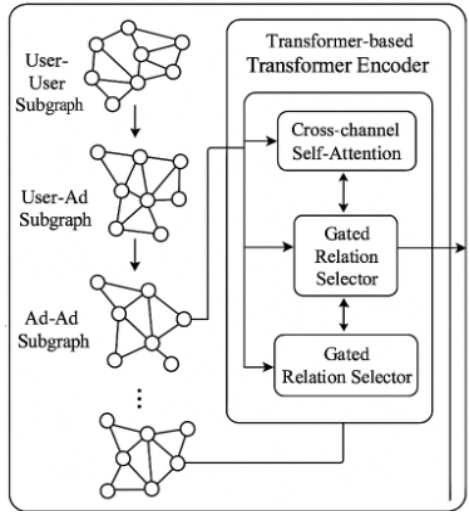

Figure 1. Hierarchical Heterogeneous Aggregation

## 4 EXPERIMENTS

### 4.1 Experimental Setup

The experiment is conducted on the Coveo Clickstream Dataset, which is the released of a real-world anonymized e-commerce clickstream dataset by Coveo and the initial SIGIR eCom data challenge dataset with over 36M user behavior events such as search, browse, click, add-to-cart and purchase, and all the user and product IDs are hashed for privacy. The statistics has multi-dimensional features including timestamps, behavior types, and product operations, and is applicable to anonymous users ad response modeling tasks. The data is available at https://github.com/coveooss/shopper-intent-prediction-nature-2020. To verify the effectiveness of the proposed model, we selected four methods related to advertising feedback prediction as comparison baseline including:

- Gated Recurrent Unit for Recommendation (GRU4Rec) feeds the behavior sequence of anonymous users into GRU4Rec to predict the class of user response to ads using final hidden state. More importantly, as a timeseries baseline without graph structure, it is useful to evaluate the gain of graph structure modeling in anonymous advertising prediction.
- Graph Attention Network (GAT) uses graph attention to construct user-ad interaction graphs, embed nodes in the graph and to predict the ad responses on the final node representations.
- Heterogeneous Graph Neural Network (HetGNN) learns the representations of nodes and relations of multiple types by leveraging structural and content information.
- Federated CTR Prediction (FedCTR) uses distributed user behavior data to build models and ensures the user privacy. FedCTR fine-tunes the model on the local device, and then synchronized the model with the weighted parameter aggregation.

### 4.2 Experimental Analysis

In the scenario of anonymous advertising feedback prediction, there are a large number of ambiguity and inter-class overlap in user behavior, and AUC is one of the core indicators that can best reflect the discriminant ability of the model. At Figure 2, we can always observe that DTH-GNN obtains highest Area Under the ROC Curve (AUC) score in every training, and also has a noticeable edge over the other four methods both in terms of convergence rate and final performance. In particular, DTH-GNN consumed little patience to penetrate AUC threshold of 0.75 at the beginning, and then consistently ascended as the rounds increase until it surpassed 0.85/30-round, whereas the strongest competitor, HetGNN, just achieved approximately 0.80 in its 30th round. Besides, DTH-GNN has the least shaded region which implies that its performance is steadier among different training batches and shows a relatively small fluctuation of performance.

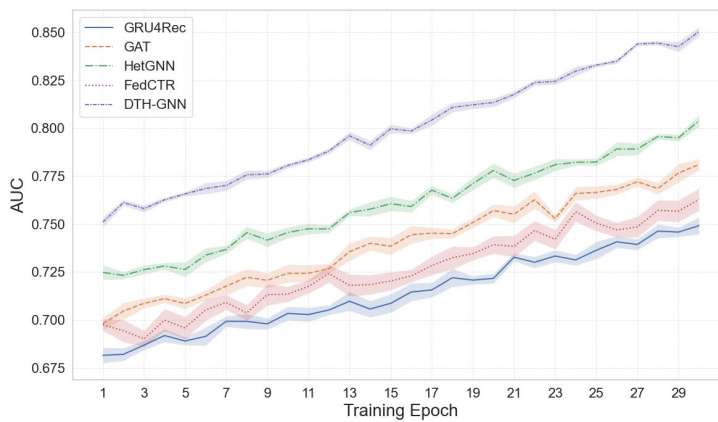

Figure 2. AUC Comparison Across Methods over Training Epochs

Specifically, the dataset includes 524,871 anonymous users and 34,602 unique ads, with a total of 36,126,487 user-ad interaction records collected over a 30-day time span. Furthermore, we define the anonymize user activity characteristics and their meanings as Table 1.

Table 1. Anonymize User Activity Characteristics and Meanings.

| Symbol | Feature | Description |
| --- | --- | --- |
| P1 | Number of pageviews | Total "pageview" events within the session |
| P2 | Number of search queries | Count of "search" operations initiated by the user in the session |
| P3 | Number of detail page views | Times the user clicked into a product detail page |
| P4 | Number of add to cart events | Count of "add-to-cart" operations performed in the session |
| P5 | Number of purchases | Total "purchase" events completed by the user |
| P6 | Session duration | Time span of the session from start to end (in seconds) |
| P7 | Average inter-event interval | Mean time interval between consecutive user actions (in seconds) |
| P8 | Unique products viewed | Number of distinct product IDs visited during the session |
| P9 | Ad impressions | Total number of ads shown to the user in the session |
| P10 | Ad clicks | Total number of ad click events executed by the user |

Subsequently, we utilize 100 Bootstrap samples, the distribution of the importance of each model to features P1–P10 was calculated for the "low activity intensity" (session events in the top 25th percentile) and "high activity intensity" (bottom 25th percentile) scenarios. It can be seen from Figure 3 that the importance distribution of features (P1–P10) is significantly different between the five methods under different activity intensities, and the boxline of DTH-GNN is largely consistent with the distribution of real data: the median and interquartile of DTH-

GNN almost overlap with the real data, especially in key features such as purchases number (P5), ad impression(P9) and clicks (P10), which is more accurate response to high-frequency interactive behaviors.

As can be seen from Table 2, DTH-GNN not only significantly outperforms the four comparison models in the two core performance indicators of AUC (91.2%) and LogLoss (0.397), but also maintains a good balance in terms of training efficiency and model complexity. Although HetGNN performed second in AUC, its training time (4.0 hours) and model parameter quantity (8.5M) were higher than those of DTH-GNN (3.9 hours, 6.8M), indicating that DTH-GNN has higher computational efficiency and lower resource occupation while capturing complex anonymous interaction patterns. In contrast, the traditional sequence model GRU4Rec and the federated model FedCTR have a small number of parameters, but the prediction accuracy is obviously insufficient.

Table 2. Performance and Efficiency Comparison of DTH-GNN and Baseline Models.

| Model | Type | AUC (%) | LogLoss | Training Time (h) | Model Size (M params) |
|---|---|---|---|---|---|
| GRU4Rec | Sequence-based | 79.3 | 0.475 | 2.1 | 4.2 |
| GAT | Graph-based (homogeneous) | 81.1 | 0.449 | 3.5 | 6.1 |
| HetGNN | Graph-based (heterogeneous) | 84 | 0.421 | 4 | 8.5 |
| FedCTR | Federated + Privacy | 82.5 | 0.438 | 4.2 | 7.2 |
| DTH-GNN (Ours) | Temporal-Hetero + Contrast | 91.2 | 0.397 | 3.9 | 6.8 |

On the other hand, other algorithms are less well suited for this because they are either too weak in high intensity, or are too localized, and fail to consider both the central tendency and the outliers of the distribution simultaneously. It can be observed that DTH-GNN not only is close to the real user behavior in the average importance level, but also preserves the diversity and tail behavior of feature distribution, which further indicates its superior performance in modeling the positive response features in an anonymous e-commerce scenario.

## 5 CONCLUSION

In conclusion, focusing on the anonymized user activity data in e-commerce platforms, we outline a DTH-GNN model that effectively models the short-term burst, diurnal cycle and long-range memory patterns with temporal edge decomposition, hierarchical heterogeneous aggregation and feedback-aware contrast regularity, dynamically fuses heterogeneous subgraphs with meta-path conditional transformer, and employs contrastive regular-stable representation. We empirically demonstrate that DTH-GNN is markedly competitive with other state-of-the-art methods. For future work, more informative multimodal features (e.g., image, text and ratings) can be adopted to diversify the model's understanding over complex commodity information, drawing on insights from multimodal interaction systems in virtual meeting and detection settings [19,20]. Additionally, fusion of structured and unstructured data, as shown in psychological health prediction using EHR for vulnerable populations [21], may inspire further anonymous user modeling in domains beyond advertising.

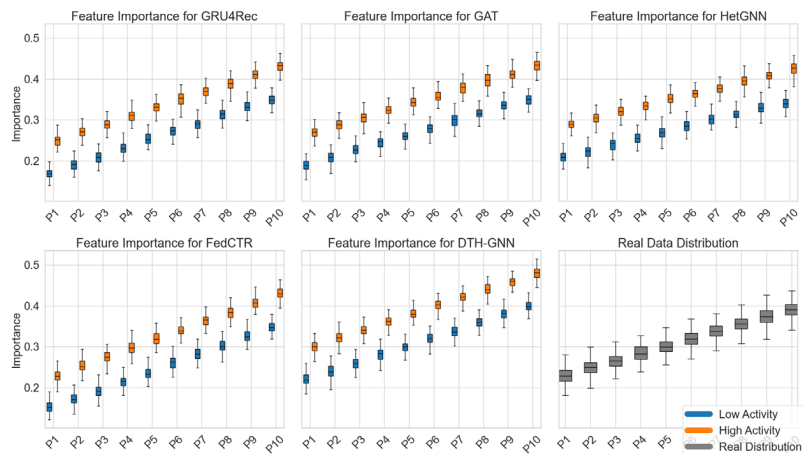

Figure 3. Feature Importance Distributions for Positive Ad Response Different Activity Intensities